\documentclass[a4paper, 11pt]{article}
\usepackage{comment} 
\usepackage[nointegrals]{wasysym}
\usepackage{fullpage} 
\usepackage{graphicx}
\usepackage{cite}
\usepackage{multicol}
\usepackage[perpage]{footmisc} 
\usepackage{sectsty}
\sectionfont{\large}
\usepackage{yhmath}
\usepackage{bm}
\usepackage{float}
\usepackage{amsmath}
\usepackage[utf8]{inputenc}
\usepackage{amssymb}
\usepackage{mathrsfs}
\usepackage{wrapfig}
\usepackage{enumitem}
\usepackage[english]{babel}
\usepackage{comment}
\usepackage{bbold}  
\usepackage[usenames,dvipsnames]{xcolor} 
\usepackage{xcolor,hyperref}
\usepackage{setspace}
\usepackage{abstract}
\usepackage{tablefootnote}

\usepackage[T1]{fontenc}
\usepackage[margin=0.94in]{geometry}
\def\Mpl{M_{\rm P}}

\usepackage{enumitem}

\begin{document}
\null\hfill 
IPMU25-0032\\ 
\null\hfill LMU-ASC 17/25\\ 
\null\hfill MPP-2025-119 \\
\vspace*{\fill}

\vspace*{\fill}
\begin{center}
    \LARGE\textbf{{ \textcolor{Black}{Conformal and pure scale-invariant gravities \\in}}} \textbf{\textit{d}} \LARGE\textbf{{dimensions} }

\end{center}

\begin{center}
  \normalsize\textsc{Anamaria Hell,$^{1}$ Dieter Lüst,$^{2,\;3}$ }
\end{center}

\begin{center}
    $^{1}$ \textit{Kavli IPMU (WPI), UTIAS,\\ The University of Tokyo,\\ Kashiwa, Chiba 277-8583, Japan}\\
    $^{2}$\textit{Arnold Sommerfeld Center for Theoretical Physics,\\
Ludwig–Maximilians–Universität München\\
Theresienstraße 37, 80333 Munich, Germany}\\
$^{3}$\textit{Max–Planck–Institut für Physik (Werner–Heisenberg–Institut)\\
Boltzmannstra{\ss}e 8, 85748 Garching, Germany}\\ 

\end{center}
\thispagestyle{empty} 

\renewcommand{\abstractname}{\textit{\textcolor{Black}{Abstract}}}

\begin{abstract}
We consider conformal and scale-invariant gravities in d dimensions, with a special focus on pure $R^2$ gravity in the scale-invariant case. In four dimensions, the structure of these theories is well known. However, in dimensions larger than four, the behavior of the modes is so far unclear. In this work, we explore this question, studying the theories in conformally flat spacetimes as well as anisotropic backgrounds. First, we consider the pure theory in d-dimensions. We show that this theory propagates no degrees of freedom for flat space-time. Otherwise, we find the theory in the corresponding Einstein frame and show that it propagates a scalar field and two tensor modes, that arise from Einstein's gravity. We then consider conformal gravity in d dimensions. We argue on the number of degrees of freedom for conformally flat space-times and show that for $d>4$, there exists a frame in which this theory can be written as the Weyl-squared gravity with a cosmological constant, and also generalize this formulation to the $f\left(W^2\right)$ theories. Then, we consider the specific model of conformal gravity in five dimensions. We find the analytical and numerical solutions for the anisotropic Universe for this case, which admits super-Hubble and exponential expansions. Finally, we consider the perturbations around these solutions and study the number of the degrees of freedom. 
  \end{abstract}
 
\vfill
\small
\noindent\href{mailto:anamaria.hell@ipmu.jp}{\text{anamaria.hell@ipmu.jp}}\\
\href{luest@mppmu.mpg.de}{luest@mppmu.mpg.de}\\
\vspace*{\fill}

\clearpage
\pagenumbering{arabic}
\newpage
\tableofcontents

\newpage

\section{Introduction}
The existence of dimensions larger than four is an intriguing possibility that might be realized in our Universe. This hypothesis is essential to several key models of high-energy and gravitational physics (see \cite{Rubakov:2001kp, Maartens:2010ar, Cheng:2010pt} for a review). Originally, it was introduced as a way to describe scalar gravity and electromagnetism \cite{Nordstrom:1914ejq}, and later proposed as a solution to the unification of electromagnetism with Einstein's gravity \cite{Kaluza:1921tu, Klein:1926tv}. The extra dimensions have also gained significant importance as a potential solution to the hierarchy problem \cite{Arkani-Hamed:1998sfv, Arkani-Hamed:1998jmv, Antoniadis:1998ig, Randall:1999ee, Randall:1999vf}, explaining the apparent gap between the weak-interactions scale and the Planck scale by making the gravitational force stronger. They can arise from string theory, one of the main theories of quantum gravity, where their existence serves as one of the consistency conditions \cite{Green:1987sp, Blumenhagen:2013fgp}. Moreover, they were essential to some of the well-known models of modified theories of gravity, that tackled the puzzle of the cosmological constant, giving rise to the self-accelerating solutions, as well as the mass to the graviton \cite{Dvali:2000hr, Deffayet:2000uy, Deffayet:2001pu, Deffayet:2001uk, Deffayet:2002sp}. 
Notably, the existence of a fifth dimension called \textit{the dark dimension}, has also gained significant interest even recently, appearing as a possible solution to the cosmological hierarchy problem -- the question of why is the dark energy so small -- when combined with the observations and swampland conjectures \cite{Lust:2019zwm,Montero:2022prj, Anchordoqui:2024ajk}, and may even give rise to the inflation \cite{Anchordoqui:2022svl, Anchordoqui:2023etp}. 

Some of the theories that naturally emerge in the scenarios with extra dimensions are higher derivative theories of gravity \cite{Alvarez-Gaume:2015rwa}. 
In four dimensions,  quadratic gravity is argued to be renormalizable and asymptotically free \cite{Stelle:1976gc, Stelle:1977ry, Julve:1978xn, Fradkin:1981iu}. 
Thus, they were considered as possible UV completions of gravity \cite{Adler:1982ri, tHooft:2011aa}. 
More recently, it was shown that higher order gravity   provides an upper bound for the ultraviolet cutoff of the effective action, since 
 the species scale  \cite{Dvali:2007hz,Dvali:2007wp,Dvali:2009ks,Dvali:2010vm,Dvali:2012uq} is controlled by the moduli dependent coefficients of higher curvature operators 
  \cite{vandeHeisteeg:2022btw,Cribiori:2022nke,Calderon-Infante:2025ldq}.
Among them, \textit{conformal gravity} (CG) is particularly interesting. This theory contains an action that is only a function of the Weyl-tensor squared, making it thus invariant under conformal transformations \cite{Weyl:1918ib, Weyl:1919fi}. As a result, it only depends on the angles, and not on the distances. At the same time, CG suffers from pathological modes -- Ostrogradsky ghost degrees of freedom, which classically lead to instability and violation of unitarity upon the quantization of the theory \cite{Ostrogradsky:1850fid}.  One should note, however, that the same issue might be removed with the appropriate choice of boundary conditions \cite{Maldacena:2011mk, Anastasiou:2016jix, Anastasiou:2020mik, Hell:2023rbf}.

Another interesting case of higher-derivative theories of gravity in four dimensions is the \textit{pure $R^2$ gravity}. This theory only contains the square of the Ricci scalar, thus making it scale-invariant \cite{Deser:2003up, Deser:2007vs, Kehagias:2015ata}. Notably, this is the only theory of quadratic gravity that is in addition to this scale invariance also ghost-free \cite{Kounnas:2014gda, Alvarez-Gaume:2015rwa}. Moreover, it has a curious structure -- in Minkowski space-time, the linearized theory has no propagating degrees of freedom, while in the cosmological background, it propagates a scalar mode together with two tensor degrees of freedom \cite{Hell:2023mph}. As shown in \cite{ Alvarez-Gaume:2015rwa}, this can be easily seen if one goes to the Einstein frame, valid for $R\neq 0$, where the theory can be written as Einstein gravity, together with a cosmological constant and a scalar field.

The higher-derivative theories of gravity can be expected to arise from the extra dimensions, in models such as string theory, supergravity, or M theory \cite{Fradkin:1985am, Alvarez-Gaume:2015rwa}. However, when emerging, they are not necessarily quadratic in curvature but can appear also in higher-polynomial and even non-polynomial combinations. It is thus curious to ask if such theories will keep the same properties as those in four dimensions. In this work, we will investigate these questions, with a particular focus on conformal and scale-invariant gravities, and ask -- \textit{How do these theories change when compared to their four-dimensional counterparts? }

Curiously, by focusing on the scale-invariant gravity, which contains only the powers of the Ricci tensor, we will show that the theory describes no degrees of freedom in the Jordan frame for the flat-spacetime, while for the non-vanishing values of the Ricci scalar, one can find the corresponding Einstein frame, which in the presence of extra dimensions admits only a positive cosmological constant. 

In contrast to the pure scale-invariant case, conformal gravity in d dimensions is significantly more complicated in d dimensions. For conformally flat spacetimes, this action is non-polynomial. Thus, it makes it more non-trivial to analyze the number of degrees of freedom. Considering general backgrounds might thus be even more demanding, due to the complicated structure of the theory. However, as pointed out in \cite{Deruelle:2009zk}, if one considers a function of the square of the Weyl-tensor, it could give rise to more degrees of freedom, in contrast to the special case of conformal gravity. Thus, we might expect that in d-dimensions, the conformal gravity has a more diverse spectrum, when compared to its four-dimensional counterpart. 

A common way to count the degrees of freedom in higher derivative gravitational theories is to find a frame in which such theories take a more simplified form. One well-known example of this procedure is the $f(R)$ gravity, which can be written as Einstein gravity coupled with a scalar field with a potential, that depends on the form of the function $f$ \cite{Teyssandier:1983zz, Whitt:1984pd, Wands:1993uu,  Sotiriou:2008rp}. One can also consider the actions for quadratic gravity, or more general functions of the Riemann and Ricci tensor, and express them in the canonical form with the help of the external fields \cite{Jakubiec:1988ef, Hindawi:1995an, Hindawi:1995cu}. Along these lines, we will show that in the case of conformal gravity, there exists an interesting frame, in which the starting action, which includes powers of the square of the Weyl-tensor, can be written in a quadratic way, together with a cosmological constant. 
While this alternative frame will hold only for space-times that are not conformally flat, it will be very useful to count the degrees of freedom. Thus, in this work, we will focus on the five-dimensional case, and find anisotropic solutions.  Moreover, by studying the perturbations around these solutions, we will show that in addition to the modes analogous to the four-dimensional case, this theory also propagates scalar modes and additional vector modes. 

The paper is organized in the following way. In section 2, we will study the pure scale invariant gravity, with a brief review of the four-dimensional case. Then, in section 3, we will consider conformal gravity, with a focus on both the conformally flat spacetimes and more general backgrounds. In section 4, we will consider the five-dimensional case, and analyze background solutions and perturbations for the theory in the alternative frame. Finally, in section 5, we will conclude and discuss our results.

\section{Scale-invariant gravity in four dimensions}
One of the main questions when studying the theories of gravity is how many degrees of freedom it describes, especially when compared to Einstein's general relativity. 
In this work, we will explore this question in the context of higher-derivative gravitational theories in d-dimensions, with a focus on pure scale-invariant and conformally invariant gravities. 

In order to understand their behavior in more depth, it will be convenient to compare them to their four-dimensional counterparts. Thus, in this section, we will first focus on degrees of freedom in the most general scale-invariant gravity four dimensions, analyzing the modes in flat-spacetime, and afterward consider the d-dimensional cases.

The most general scale invariant gravity in four dimensions is described by the following action:
\begin{equation}\label{quadratic_gravity_4}
    S=\int d^4x\sqrt{-g}\left[\alpha_{CG} R_{\mu\nu} R^{\mu\nu}+\beta R^2 + \gamma R_{\mu\nu\rho\sigma}R^{\mu\nu\rho\sigma}+\lambda R_{\mu\nu\rho\sigma} \Tilde{R}^{\mu\nu\rho\sigma}\right],
\end{equation}
where
\begin{equation}
    \Tilde{R}^{\mu\nu\rho\sigma}=\frac{1}{2}\varepsilon^{\mu\nu\alpha\beta}R_{\alpha\beta}^{\;\;\;\rho\sigma}
\end{equation}
is the dual, of the Ricci tensor. In four dimensions, the third term can be replaced with the Gauss-Bonnet term, which is a total derivative:
\begin{equation}
    \textit{GB}=R_{\mu\nu\rho\sigma}R^{\mu\nu\rho\sigma}-4R_{\mu\nu} R^{\mu\nu}+R^2,
\end{equation}
although, one should note that in a higher number of dimensions, this no longer holds. 
The last term is a topological one as well. Therefore, the dynamics of scale-invariant gravity are mainly encoded only in the first two terms. In the following, we will review their behavior in four dimensions. 

Let us first consider the second term in the action (\ref{quadratic_gravity_4}). We will perturb the metric in the following way:
\begin{equation}
    g_{\mu\nu}=\eta_{\mu\nu}+h_{\mu\nu},
\end{equation}
and decompose its perturbations into the irreducible representation of the group of rotations as:
\begin{equation}
    \begin{split}
        h_{00}&=2\phi\\
        h_{0i}&=S_i+B_{,i}\\
        h_{ij}&=2\psi\delta_{ij}+E_{,ij}+F_{i,j}+F_{j,i}+h^T_{ij}, 
    \end{split}
\end{equation}
where the comma $_{,i}$ denotes a derivative with respect to the coordinate $x^i$, and vector and tensor modes satisfy:
\begin{equation}
    S_{i,i}=0\qquad F_{i,i}=0\qquad h_{ij,i}=0\qquad h_{ii}=0. 
\end{equation}
Then, in the conformal and Poisson gauge
\begin{equation}
    E=0 \qquad\qquad B=0\qquad\qquad F_i=0
\end{equation}
in which the scalar and vector perturbations coincide with the gauge-invariant ones for the linearized theory \cite{Bardeen:1980kt, Mukhanov:2005sc}, one finds the following Lagrangian density:
\begin{equation}
    \mathcal{L}=4\beta(\Delta \phi+3\Ddot{\psi}-2\Delta\psi)^2.
\end{equation}
We can see that $\phi$ is a constrained field, that satisfies:
\begin{equation}
    \phi=-\frac{1}{\Delta}\left(\Ddot{\psi}-2\Delta\psi\right)
\end{equation}
By substituting this back to the action, one finds that the Lagrangian density vanishes,
\begin{equation}
    \mathcal{L}_S=0.
\end{equation}
Thus, if the space-time is flat, the second term in the action (\ref{quadratic_gravity_4}) describes no degrees of freedom \cite{Hell:2023mph}. 

Let us now consider the first term in (\ref{quadratic_gravity_4}). In this case, for flat space-time, we find the following Lagrangian density: 
\begin{equation}
    \begin{split}
        \mathcal{L}&=\mathcal{L}_S+\mathcal{L}_V+\mathcal{L}_T\\
        \mathcal{L}_S&=\frac{2\alpha_{CG}}{3}\left(\phi\Delta^2\phi+2\phi\Delta^2\psi+\psi\Delta^2\psi\right)^2-\frac{8\alpha_{CG}}{3}(\Delta \phi+3\Ddot{\psi}-2\Delta\psi)^2\\
        \mathcal{L}_V&=\frac{\alpha_{CG}}{2}\left(\Ddot{S}_i\Delta S_i-S_i\Delta^2S_i\right)\\
        \mathcal{L}_T&=\frac{\alpha_{CG}}{4}\Box h_{ij}^T\Box h_{ij}^T
    \end{split}   
\end{equation}
In contrast to the previous case, we can notice that this theory includes two tensor modes, one of which is a ghost, and a healthy vector mode. Similarly to the previous procedure, let us find only the propagating modes. We can notice that the scalar mode $\phi$ is again not propagating. By varying the action with respect to it, we find the following constraint: 
\begin{equation}
    \Delta\phi=3\Delta\psi-4\Ddot{\psi}.
\end{equation}
By substituting its solution back to the action, the Lagrangian density describing the scalar modes becomes:
\begin{equation}
    \mathcal{L}_S=8\alpha_{CG}\Box\psi\Box\psi.
\end{equation}
Thus, we find two scalar degrees of freedom -- a healthy mode and a ghost one. However, by choosing a particular combination of the $\alpha_{CG}$ and $\beta$ couplings, 
\begin{equation}
    \beta=-\frac{1}{3}\alpha_{CG}\qquad\text{and}\qquad \alpha_{CG}=2\alpha_{CG},
\end{equation}
the above action becomes the action for conformal gravity, up to the Gauss-Bonnet term\footnote{Note that the Gauss-Bonnet term gives a non-vanishing contribution in the case of the FLRW metric, and is thus crucial addition to the combination of Ricci scalar and Ricci tensor which makes it conformally invariant \cite{Hell:2023rbf}. }:
\begin{equation}
    S_{CG}=\alpha_{CG}\int d^4x\sqrt{-g}\left(2R_{\mu\nu}R^{\mu\nu}-\frac{2}{3}R^2+\text{GB}\right).
\end{equation}
This theory in conformally flat spacetimes retains the same structure for the vector and tensor modes as $R_{\mu\nu}R^{\mu\nu}$ gravity. However, the scalar sector now stops to propagate. In particular, we can see that the Lagrangian density corresponding to it is given by:
\begin{equation}
    \mathcal{L}=\mathcal{L}_{S,CG}=\frac{4\alpha_{CG}}{3}\left(\phi\Delta^2\phi+2\phi\Delta^2\psi+\psi\Delta^2\psi\right)^2. 
\end{equation}
Thus, it yields the constraint:
\begin{equation}
    \phi+\psi=0,
\end{equation}
which sets the scalar modes to zero. 

Therefore, so far, we have seen that higher-derivative gravity brings a rich spectrum when compared to Einstein's Relativity. In particular, for the flat space-time case, we have confirmed that the above action in general has pathological degrees of freedom -- one ghost in the tensor sector, and one in the scalar one. In addition to those, the general theory also has a healthy scalar mode, a healthy tensor one, and two well-behaved vector modes \cite{Stelle:1976gc, Julve:1978xn, Alvarez-Gaume:2015rwa}. If one sets the coupling constants between different terms such that the action is conformally invariant, the scalar sector vanishes, and the resulting conformal gravity only has two vector modes, and four tensor degrees of freedom, among which one is pathological. However, curiously, the pure $R^2$ gravity in flat space-time has no propagating modes in any sector for flat space-time. In contrast, if one considers the curved space-time, it propagates a tensor dof and a scalar mode \cite{Alvarez-Gaume:2015rwa, Hell:2023mph}. Let us now see how this pure case and a conformally invariant one generalize when we instead consider the theory in d-dimensions.

\section{The pure scale-invariant gravity in d-dimensions}

In the previous section, we have seen that the scale-invariant gravity in four dimensions involves a combination of only four terms, among which two give rise to propagating modes -- the squares of the Ricci scalar and the Ricci tensor. The remaining two terms that contain the Riemann tensor, on the other hand, are topological.

In the case of d-dimensions, however, one can have much more possibilities, with examples that include:
\begin{equation}
    R^{\frac{d}{2}},\qquad (R_{\mu\nu}R^{\mu\nu})^{\frac{d}{4}},\qquad (\text{GB})^{\frac{d}{4}}, \qquad (RR_{\mu\nu}R^{\mu\nu})^{\frac{d}{6}},\qquad  (R^2R_{\mu\nu\rho\sigma}R^{\mu\nu\rho\sigma})^{\frac{d}{8}}
\end{equation}
In this work, we will be interested in the simplest case among those, that involves only the powers of the Ricci scalar. In particular, one can find that the action for the pure, scale-invariant gravity in d-dimensions is given by: 
\begin{equation}
    S_{pure}=\beta_d \int d^dx R^{\frac{d}{2}}
\end{equation}
Here, $\beta_d$ is the dimensionless coupling constant. In the following, we will study the degrees of freedom of this theory in flat and curved space-time. 

\subsection{The flat space-time}
Let us first consider the pure scale-invariant gravity in flat space-time. In contrast to the four-dimensional case, this theory is non-linear from the start. By perturbing the metric around flat space-time, 
\begin{equation}
    g_{\mu\nu}=\eta_{\mu\nu}+h_{\mu\nu}
\end{equation}
we find
\begin{equation}
    S_{pure}=\beta_d\int d^dx (\partial^{\mu}\partial^{\nu}h_{\mu\nu}-\Box h)^{\frac{d}{2}} 
\end{equation}
Following the approach of the previous section, we can further examine the degrees of freedom by decomposing the metric perturbations into the scalar, vector, and tensor modes: 
\begin{equation}\label{ddimdec}
    \begin{split}
        h_{00}&=2\phi\\
        h_{0i}&=S_i+B_{,i}\\
        h_{ij}&=2\psi\delta_{ij}+E_{,ij}+F_{i,j}+F_{j,i}+h^T_{ij}, 
    \end{split}
\end{equation}
where now $i=1,...,d-1$, and vector and tensor modes satisfy:
\begin{equation}
    S_{i,i}=0\qquad F_{i,i}=0\qquad h_{ij,i}=0\qquad h_{ii}=0. 
\end{equation}
We can notice that the vector and tensor modes drop out from the above action. What remains are the scalar modes, whose action in the d-dimensional conformal gauge 
\begin{equation}
    E=0\qquad\qquad B=0,
\end{equation}
given by
\begin{equation}
    S_{pure}=\int d^dx 2^{d/2}(\Delta\phi+\Ddot{\psi}-2\Delta\psi)^{d/2}
\end{equation}
Similarly to the four dimensions, the scalar mode $\phi$ is not propagating. By varying the action with respect to it, we find that it satisfies the following constraint:
\begin{equation}
    \Delta\left[\left(\Delta\phi+\Ddot{\psi}-2\Delta\psi \right)^{\frac{d-2}{d}}\right]=0,
\end{equation}
whose solution is given by: 
\begin{equation}
    \phi=-\frac{1}{\Delta}(\Ddot{\psi}-2\Delta\psi). 
\end{equation}
By substituting it back to the action, we find: 
\begin{equation}
    S_{pure}=0. 
\end{equation}
Thus, similarly to the four-dimensional case, the leading order theory in flat space-time for pure scale-invariant gravity does not have any propagating degrees of freedom.

\subsection{The curved space-time and the Einstein frame}
As a next step, let us consider an arbitrary background, and count the number of propagating modes. In this case, it is much more convenient to go to the Einstein frame. In order to do this, we first introduce a scalar field: 
\begin{equation}
    S=\frac{d(d-2)}{4}\int d^dx\sqrt{-g}\left[\frac{2}{d-2}R\phi^{\frac{d-2}{2}}-\frac{2}{d}\beta_d^{\frac{-2}{d-2}}\phi^{\frac{d}{2}}\right]
\end{equation}
This field is constrained. By varying the action with respect to it, we find:
\begin{equation}
    \phi=\beta_d^{\frac{2}{d-2}}R,
\end{equation}
and thus, by substituting this back to the action, we recover the initial action for the pure scale invariant gravity:
\begin{equation}
    S=\beta_d\int d^dx\sqrt{g}R^{\frac{d}{2}}. 
\end{equation}
As a next step, we define a new metric:   
\begin{equation}
    \Tilde{g}_{\mu\nu}=\frac{\phi}{M_P^2}g_{\mu\nu}, 
\end{equation}
and express the original one in terms of it. In this case, the Ricci scalar can be written as: 
\begin{equation}
    R=F\Tilde{R}+(d-1)\Tilde{\nabla}_{\mu}\Tilde{\nabla}^{\mu}F-\frac{(d-1)(d+2)}{4}\frac{\Tilde{\nabla}_{\mu}F\Tilde{\nabla}^{\mu}F}{F}
\end{equation}
where
\begin{equation}
    F=\frac{\phi}{M_P^2},
\end{equation}
$\Tilde{R}$ is the Ricci scalar corresponding to the new metric, and $\Tilde{\nabla}_{\mu}$ the covariant derivative associated with the new metric. By substituting this into action, and performing integration by parts, we find:
\begin{equation}\label{pureRGen}
    S=\int d^dx\sqrt{-\Tilde{g}}\left[\frac{d M_P^2}{2}\Tilde{R}-\frac{d(d-2)(d-1)}{8M_P^2\phi^2}\Tilde{\nabla}_{\mu}\phi\Tilde{\nabla}^{\mu}\phi-M_P^{d-2}\Lambda\right]
\end{equation}
where
\begin{equation}
    \Lambda=\frac{d-2}{2}M_P^2\beta_d^{\frac{-2}{d-2}}
\end{equation}
is the cosmological constant. Therefore, similarly to the pure case in four dimensions, we have arrived at an action for Einstein's gravity, coupled with a scalar field, and a cosmological constant. However, in contrast to the four-dimensional case, we can see that this cosmological constant can only be positive. The reason for this lies in the coupling constant. In general, it can be positive or negative, and thus in four dimensions, it affects the overall signature of $\Lambda$. However, in the presence of extra dimensions, the coupling is always squared, thus resulting only in positive values of the cosmological constant. 

The scalar field above can be further canonically normalized by defining 
\begin{equation}
    \Phi=\frac{\sqrt{d(d-1)(d-2)}}{2M_P}\ln\phi. 
\end{equation}
Then, the action (\ref{pureRGen}) becomes:
\begin{equation}
    S=\int d^dx\sqrt{-\Tilde{g}}\left[\frac{d M_P^2}{2}\Tilde{R}-\frac{1}{2}\Tilde{\nabla}_{\mu}\Phi\Tilde{\nabla}^{\mu}\Phi-dM_P^{d-2}\beta_d^{\frac{-2}{d-2}}\Lambda\right]
\end{equation}

Therefore, in the general case, we find that the scale-invariant gravity is equivalent to d-dimensional Einstein gravity with a cosmological constant and a free scalar field. One should note, however, that the above transformation holds only for $R\neq 0$, similarly to the four-dimensional case. If this holds, then one should consider the pure scale-invariant gravity in the Jordan frame, in which there are no degrees of freedom propagating for flat space-time.

\section{Conformal gravity in d-dimensions}
Previously, we have seen that pure scale-invariant gravity has results that are analogous to the four-dimensional case, up to the sign of the cosmological constant. Let us now see if the same will hold if one considers conformal gravity. In four dimensions, it is given by the following action:
\begin{equation}\label{CG44}
    S=\int d^4x\sqrt{-g}W_{\mu\nu\rho\sigma}W^{\mu\nu\rho\sigma}
\end{equation}
where $W_{\mu\nu\rho\sigma}$ is the Weyl tensor, which is defined by:
\begin{equation}
    W_{\mu\nu\rho\sigma}= R_{\mu\nu\rho\sigma}-\frac{1}{d-2}\left(g_{\mu\rho}R_{\nu\sigma}-g_{\mu\sigma}R_{\nu\rho}-g_{\nu\rho}R_{\mu\sigma}+g_{\nu\sigma}R_{\mu\rho}\right)+\frac{1}{(d-1)(
    d-2)}R\left(g_{\mu\rho}g_{\nu\sigma}-g_{\mu\sigma}g_{\nu\rho}\right)
\end{equation}
The action (\ref{CG44}) describes a healthy vector mode and two tensor ones, one of which is a ghost when the space-time is conformally flat space-times. Moreover, it is invariant under conformal transformations:
\begin{equation}
    g_{\mu\nu}\to\Tilde{g}_{\mu\nu}=\Omega^2(x) g_{\mu\nu}.
\end{equation}
This holds because the quantities in the action transform as 
\begin{equation}
    \sqrt{-\Tilde{g}}=\Omega^4\sqrt{-g}\qquad\text{and}\qquad \Tilde{W}_{\mu\nu\rho\sigma}\Tilde{W}^{\mu\nu\rho\sigma}=\frac{1}{\Omega^4}W_{\mu\nu\rho\sigma}W^{\mu\nu\rho\sigma}.
\end{equation}
However, similarly to the scale-invariant gravity, in d dimensions, the previous action is no longer conformally invariant. In this case, the square of the Weyl tensor transforms as:
\begin{equation}
    \Tilde{W}_{\mu\nu\rho\sigma}\Tilde{W}^{\mu\nu\rho\sigma}=\frac{1}{\Omega^4}W_{\mu\nu\rho\sigma}W^{\mu\nu\rho\sigma}
\end{equation}
However, for the metric determinant, we now find: 
\begin{equation}
  \sqrt{-\Tilde{g}}=\Omega^{d}\sqrt{-g}. 
\end{equation}
To find the action for conformal gravity in d-dimensions, we combine the above two transformation rules, and find the following expression: 
\begin{equation}\label{CGdim}
    S=\alpha_{CG}\int d^dx\sqrt{-g}\left[W_{\mu\nu\rho\sigma}W^{\mu\nu\rho\sigma}\right]^{\frac{d}{4}}
\end{equation}

As pointed out in \cite{Fradkin:1985am,Deser:1993yx,Metsaev:2010kp,  Lu:2011ks}, one can also form more complicated contractions of the 
Weyl tensor, such as, for instance, 
\begin{equation}
    W_{\mu\nu\rho\sigma}W^{\mu\nu\alpha\beta}W_{\alpha\beta}^{\;\;\;\;\rho\sigma},
\end{equation}
and based on such contractions, or more complicated ones, possibly form conformal terms as well. However, in this work, we will focus only on the simplest case. In particular, in the next subsection, we will consider the number of degrees of freedom for conformally flat space-time.

\subsection{Formulation in another frame}
As a first step, let us consider the action (\ref{CGdim}) in the conformally flat background, given by:
\begin{equation}
    ds^2=a^2(\eta)\eta_{\mu\nu}dx^{\mu}dx^{\nu}
\end{equation}
where $\eta_{\mu\nu}$ is the d-dimensional Minkowski metric. 
For such space-times the Weyl tensor vanishes, and thus, we can decompose the metric perturbations into the scalar, vector, and tensor modes, as in (\ref{ddimdec}). 

It is convenient to first find the form of the square of the Weyl tensor. In contrast to the four-dimensional case, the scalar, vector, and tensor modes do not decouple. Thus, we find in the leading order:
\begin{equation}
    \mathcal{L}=\alpha_{CG}\left[W^2\right]^{\frac{d}{4}}
\end{equation}
where
\begin{equation}
    W^2=P_S+P_V+P_T+P_{SV}+P_{ST}+P_{T}
\end{equation}
\begin{equation*}
    \begin{split}
      P_S&=\frac{4(d-3)}{d-2}\chi_{,ij}\chi_{,ij}-\frac{4(d-3)}{(d-1)(d-2)}\Delta\chi\Delta\chi\\
      P_V&=\frac{2(d-3)}{d-2}(\dot{S}_{i,j}\dot{S}_{i,j}+\dot{S}_{i,j}\dot{S}_{j,i})+2S_{i,jk}S_{k,ij}-2S_{k,ij}S_{k,ij}+\frac{2}{d-2}\Delta S_i\Delta S_i\\
      P_T&=h_{ij,\mu\nu}^Th_{ij}^{T,\mu\nu}-\frac{1}{d-2}\Box h_{ij}^T\Box h_{ij}^T+2(\dot{h}_{ij,k}^T\dot{h}_{jk,i}-h_{ij,kl}^Th_{jl,ik}^T)+h_{ij,kl}^Th_{kl,ij}^T\\
      P_{SV}&=-\frac{8(d-3)}{d-2}\dot{S}_{j,i}\chi_{,ij}\\
      P_{ST}&=\frac{4(d-3)}{d-2}\Ddot{h}_{ij}^T\chi_{,ij}+\frac{4}{d-2}\Delta h_{ij}^T\chi_{,ij}\\
      P_{VT}&=-2(\dot{S}_{i,j}\Ddot{h}_{ij}^T-S_{i,kj}\dot{h}_{ik,j}+\dot{S}_{j,i}\Ddot{h}_{ij}^T+S_{i,kl}\dot{h}_{kl,i}^T-2S_{k,ij}\dot{h}_{ij,k}^T)-\frac{4}{d-2}\dot{S}_{j,i}\Box h_{ij}^T
    \end{split}
\end{equation*}
Here, the dot stands for the derivative with respect to the conformal time. 

In flat space-time with d dimensions and a quadratic Weyl tensor, the action given by
\begin{equation}
    S=\int d^dx\sqrt{-g}W_{\mu\nu\rho\sigma}W^{\mu\nu\rho\sigma}=\int d^dx\left(P_S+P_V+P_T\right).
\end{equation}
Thus, the theory propagates the same types of modes as in conformal gravity in four dimensions, and with all modes decoupled. In particular, the scalar dof does not propagate and is rather constrained, while the vector degree of freedom and two tensor modes propagate, with one of them being a ghost.

In the conformally-invariant case, in contrast, the modes are not decoupled.  However, similarly to the theory in four dimensions, we can notice that the scalar modes are again not propagating. In particular, they satisfy the following constraints:
\begin{equation}
    \begin{split}
        \frac{8(d-3)}{(d-2)}\partial_i\partial_j\left(f\chi_{,ij}-f\dot{S}_{j,i}-\frac{1}{2}f\Ddot{h}_{ij}^T\right)+\frac{4}{d-2}\partial_i\partial_j(f\Delta h_{ij}^T)-  \frac{8(d-3)}{(d-2)(d-1)}\Delta(f\Delta\chi)=0
    \end{split}
\end{equation}
where 
\begin{equation}
    f=(W_{\mu\nu\rho\sigma}W^{\mu\nu\rho\sigma})^{\frac{d-4}{4}}
\end{equation}
While we cannot solve the above constraint exactly due to the coupling between different modes, we can notice that the largest number of derivatives that appear in the above constraint is two, arising from the tensor perturbations. This indicates to us that there will be no higher than four-time derivatives in this leading order of the perturbation theory. Moreover, the vector modes are propagating, with a form that suggests only one mode present in the theory. Thus, while we have not proven it exactly, the form of the above Lagrangian density indicates that there will be a single vector mode and two tensors, while the scalars remain absent.

The above non-polynomial form of the action is nevertheless non-trivial, making it thus complicated to study the number of propagating modes. In order to simplify this procedure, it is thus natural to ask if there is another formulation of the theory, which would make the apparent difficulty absent. We will explore this in the following subsection.

\subsection{An alternative frame}
In the previous case, we have considered degrees of freedom in conformal gravity for the conformally flat metric. In this subsection, we will focus on a more general case. In particular, our goal is to write the action (\ref{CGdim}) in a way that allows us to read off the degrees of freedom in a more clear way. For this, let us introduce a scalar field in the following way:
\begin{equation}
    S=\alpha_{CG}\int d^dx\sqrt{-g}(W^2)^{d/4}=\alpha_{CG}\int d^dx\sqrt{-g}\left[\frac{d}{4}W_{\mu\nu\rho\sigma}W^{\mu\nu\rho\sigma}\phi^{\frac{d-4}{4}}-\frac{d-4}{4}\phi^{\frac{d}{4}}\right]
\end{equation}
By varying the action with respect to the scalar, we find:
\begin{equation}
    \phi=W_{\mu\nu\rho\sigma}W^{\mu\nu\rho\sigma}
\end{equation}
Thus, by substituting this back into the above action, we recover the original one. As a next step, let us define a metric: 
\begin{equation}
    \Tilde{g}_{\mu\nu}=\frac{\sqrt{\phi}}{M_P^2}g_{\mu\nu}. 
\end{equation}
Then, by writing the action in terms of it, we find: 
\begin{equation}
    S=\int d^dx \sqrt{-\Tilde{g}}\left[\frac{\alpha_{CG}dM_P^{d-4}d}{4}\Tilde{W}_{\mu\nu\rho\sigma}\Tilde{W}^{\mu\nu\rho\sigma}-M_P^{d-2}\Lambda\right]
\end{equation}
where 
\begin{equation}
    \Lambda=M_P^2\alpha_{CG}\frac{d-4}{4}
\end{equation}
is the cosmological constant. Clearly, it vanishes in four dimensions, where the previous action corresponds to the standard conformally invariant action with the Weyl-squared gravity. 

One should note that the above metric is conformally invariant. This could be exciting from the point of view of the Neumann or ABC conditions in four dimensions, as a way of removing the ghost dof and recovering Einstein gravity \cite{Maldacena:2011mk, Hell:2023rbf}. In four dimensions, such conditions break conformal invariance explicitly. However, if one would impose it for this action, the conformal invariance would still be preserved in the theory. 

The above frame is an analog of the Einstein frame for the $f(R)$ gravity, performed for the conformal one. Thus, one might naturally wonder if it is possible to perform the above transition also for a more general function of the Weyl tensor. We will explore this in the following subsection. 

\subsection{Extension to the $f(W^2)$ theory}

In the previous subsection, we have made a transition from the original frame of conformal gravity to the one, where the Weyl tensor appears in the bilinear form. 
We can also extend this procedure to an arbitrary function of the square of the Weyl tensor: 
\begin{equation}
    S=\int d^dx f(W^2).
\end{equation}
For this, we introduce the scalar field in the following way:
\begin{equation}
    S=\alpha_{CG}\int d^dx\sqrt{-g}\left(f(\phi)-f'(\phi)(\phi-W^2)\right)
\end{equation}
which yields a constraint:
\begin{equation}
    \phi=W^2,
\end{equation}
provided that $f''(\phi)\neq 0$. Then, by performing a conformal transformation: 
\begin{equation}
    g_{\mu\nu}=\frac{1}{F}\Tilde{g}_{\mu\nu},\qquad F=\left(\frac{4f'(\phi)}{d\Mpl^{d-4}}\right)^{\frac{2}{d-4}}
\end{equation}
we find:
\begin{equation}
    S=\alpha_{CG}\int d^dx \left[\frac{d \Mpl^{d-4}}{4}\Tilde{W}_{\mu\nu\rho\sigma}\Tilde{W}^{\mu\nu\rho\sigma}+\left(f(\phi)-\phi f'(\phi)\right)\left(\frac{4f'(\phi)}{d\Mpl^{d-4}}\right)^{\frac{-d}{d-4}}\right]
\end{equation}

Thus, we find that more general functions of the square of the Weyl tensor can be written in the form bilinear in the Weyld tensor, and coupled to the scalar field. In contrast to the $f(R)$ case, however, the scalar in the above theory is constrained, appearing without a kinetic term.

\section{Specific model: Conformal gravity in five dimensions } 
Having an alternative formulation of the theory allows us to study the degrees of freedom in a simpler way, as the action now quadratic in the Weyl tensor: 
\begin{equation}
    S=\int d^dx \sqrt{-\Tilde{g}}\left[\frac{d\alpha_{CG}M_P^{d-4}d}{4}\Tilde{W}_{\mu\nu\rho\sigma}\Tilde{W}^{\mu\nu\rho\sigma}-M_P^{d-2}\Lambda\right]
\end{equation}
However, one should note that the above solution does not hold for the conformally flat metric, because the Weyl tensor in this case vanishes, making thus the transformation singular. In this section, we will thus explore the theory in this new frame, focusing on the anisotropic background in five dimensions. For simplicity, let us consider the following metric:
\begin{equation}\label{anis5}
    ds^2=-N^2(t)dt^2+a^2(t)\delta_{ij}dx^idx^j+b^2(t)dl^2.
\end{equation}
Here, $N$ is the lapse, $a$ and $b$ are the scale factors that only depend on time, and the index i has values 1, 2 and 3. In the following, we will study the solutions for the above metric, and after this perform the perturbation theory to find out the number of the propagating modes. 

\subsection{The background solutions}
By substituting the background (\ref{anis5}) into the action, and varying with respect to the lapse, we find the constraint equation: 
\begin{equation}\label{5dconstraint}
    \begin{split}
        0&=\left(-5 \Mpl \dot{b} a^{3}   +5 \Mpl b a^{2} \dot{a}   \right)\dddot{a}-\frac{5 \Mpl b a^{2} {\ddot{a}}^{2}  }{2}-\frac{5 \Mpl\ddot{b} \dot{b}^{2} a^{4}  }{b^{2}}-\frac{5 \Mpl a^{4} {\ddot{b}}^{2}  }{2 b}-\frac{15 \Mpl b \dot{a}^{4}  }{2}\\&+\left(-15 \Mpl\ddot{b} a^{2}   -\frac{35 \Mpl    a^{2} \dot{b}^{2}}{2 b}\right) \dot{a}^{2}+\left(5 \Mpl\ddot{b} a^{3}   -\frac{5 \Mpl \dot{b}^{2} a^{3}  }{b}+5 \Mpl b a \dot{a}^{2}   \right)\ddot{a}\\&+\frac{5 \Mpl\dddot{b} \dot{b} a^{4}  }{b}-\frac{\Mpl^{3} b a^{4}}{\alpha_{CG}} \Lambda+\left(-5 \Mpl\dddot{b} a^{3}   +\frac{5 \Mpl \dot{b}^{3} a^{3}  }{b^{2}}+\frac{20 \Mpl\ddot{b} \dot{b} a^{3}  }{b}\right) \dot{a}+20 \Mpl    a \dot{a}^{3} \dot{b}
    \end{split}
\end{equation}
Furthermore, by varying with respect to the first scale factor, $a(t)$, we find:
\begin{equation}\label{5deoma}
    \begin{split}
       0&= 5 \Mpl a^{4} b\ddddot{a}   +\left(10 \Mpl a^{4} \dot{b}   +10 \Mpl a^{3} b \dot{a}   \right)\dddot{a}+\frac{15 \Mpl a^{3} b {\ddot{a}}^{2}  }{2}-10 \Mpl a^{2} \dot{b} \dot{a}^{3}  \\&+\left(-30 \Mpl a^{2} b \dot{a}^{2}   -20 \Mpl a^{4}\ddot{b}   -\frac{5 \Mpl a^{4} \dot{b}^{2}  }{b}+45 \Mpl a^{3} \dot{b} \dot{a}   \right)\ddot{a}+\frac{15 \Mpl a b \dot{a}^{4}  }{2}\\&-\frac{5 \Mpl a^{5}\ddot{b} \dot{b}^{2}  }{b^{2}}-5 \Mpl a^{5}\ddddot{b}   +\frac{25 \Mpl a^{5} {\ddot{b}}^{2}  }{2 b} +\left(-\frac{5 \Mpl \dot{b}^{2} a^{3}  }{2 b}+5 \Mpl\ddot{b} a^{3}   \right) \dot{a}^{2}\\&+\left(-25 \Mpl a^{4}\dddot{b}   +\frac{5 \Mpl a^{4} \dot{b}^{3}  }{b^{2}}-\frac{10 \Mpl a^{4}\ddot{b} \dot{b}  }{b}\right) \dot{a}-\frac{3 \Mpl^{3} a^{5} b}{\alpha_{CG}} \Lambda +\frac{5 \Mpl a^{5}\dddot{b} \dot{b}  }{b}
    \end{split}
\end{equation}
and finally, by varying with respect to the last scale factor, $b(t)$, we find:
\begin{equation}\label{5deomb}
    \begin{split}
       0&=-5 \Mpl a^{5}\ddddot{a}   +\left(-\frac{5 \Mpl a^{5} \dot{b}  }{b}-15 \Mpl a^{4} \dot{a}   \right)\dddot{a}-\frac{5 \Mpl a^{4} {\ddot{a}}^{2}  }{2}-\frac{10 \Mpl a^{6}\dddot{b} \dot{b}  }{b^{2}}\\&+\left(20 \Mpl a^{3} \dot{a}^{2}   +\frac{10 \Mpl a^{5}\ddot{b}  }{b}+\frac{10 \Mpl a^{5} \dot{b}^{2}  }{b^{2}}-\frac{40 \Mpl a^{4} \dot{b} \dot{a}  }{b}\right)\ddot{a}+\frac{10 \Mpl a^{6}\ddot{b} \dot{b}^{2}  }{b^{3}}\\&+\frac{5 \Mpl a^{6}\ddddot{b}  }{b}+\left(\frac{45 \Mpl \dot{b}^{2} a^{4}  }{2 b^{2}}+\frac{15 \Mpl a^{4}\ddot{b}  }{b}\right) \dot{a}^{2}+\frac{5 \Mpl a^{2} \dot{a}^{4}  }{2}-\frac{15 \Mpl a^{6} {\ddot{b}}^{2}  }{2 b^{2}}\\&+\left(\frac{30 \Mpl a^{5}\dddot{b}  }{b}-\frac{10 \Mpl a^{5} \dot{b}^{3}  }{b^{3}}-\frac{15 \Mpl a^{5}\ddot{b} \dot{b}  }{b^{2}}\right) \dot{a}-\frac{15 \Mpl a^{3} \dot{b} \dot{a}^{3}  }{b}-\frac{\Mpl^{3} a^{6} }{\alpha_{CG}}\Lambda
    \end{split}
\end{equation}

In the following, we will explore analytical and numerical solutions for this model. 

\subsubsection{Analytical solutions}
In order to find the analytical solutions of the model, we will consider two simplified cases, when one of the scale factors is set to unity. 

\begin{center}

{\textit{Case 1: $b(t)=1$ }} 

\end{center}
 First, let's set the scale factor $b(t)$ to unity. Then, we can add the equations of motion for $a$ and $b$ in such a way that the fourth-order derivatives for the scale factor $a(t)$ cancel. In particular, we find:
\begin{equation}
    \begin{split}
        \frac{12 \Mpl^{2} a^{4}}{\alpha_{CG}} \Lambda +15 a^{2} \dddot{a} \dot{a}   -15 a^{2} {\ddot{a}}^{2}   +30 a \ddot{a} \dot{a}^{2}   -30 \dot{a}^{4}  =0
    \end{split}
\end{equation}
Furthermore, we can eliminate the third order derivative, by solving the constraint equation for it, and substituting into the above equation. Then, the equation of motion for the scale factor becomes: 
\begin{equation}
    \frac{\Mpl^{2} a^{4}}{\alpha_{CG}} \Lambda -\frac{ a^{2} {\ddot{a}}^{2}  }{2}+ a \ddot{a} \dot{a}^{2}   -\frac{ \dot{a}^{4}  }{2}=0
\end{equation}
and admits exact solutions:
\begin{equation}
    a(t)=c_1 \exp\left({\frac{\Mpl\sqrt{\Lambda}}{\sqrt{2\alpha_{CG}}}t^2-c_2t}\right)\qquad\text{and}\qquad a(t)=c_3 \exp\left({-\frac{\Mpl\sqrt{\Lambda}}{\sqrt{2\alpha_{CG}}}t^2-c_4t}\right)
\end{equation}
In the above relations, the constants $c_1, c_2, c_3$ and $c_4$ are the constants of integration. Thus,  first corresponds to the super-Hubble expansion, together with an exponential one, which dominates at $t=0$, depending on the sign of the integration constant. The second can either be fully decaying, or exponentially expanding depending also on the sign of the integration constant.  

\begin{center}

{\textit{Case 2: $a(t)=1$ }} 

\end{center}

Let us now consider the second case, in which we will set $a(t)=1$. Then, we can add the equations for the scale factors in such a way that now the fourth-order derivative of $b$ cancels, finding the following expression:
\begin{equation}
     \frac{ -12 \Mpl^{2}}{\alpha_{CG}} \Lambda b^{3}-15 \dddot{b} \dot{b}    b+15 {\ddot{b}}^{2}    b+15 \ddot{b} \dot{b}^{2}  =0
\end{equation}
By using the constraint equation to express the third-order derivative in terms of the other terms, we find: 
\begin{equation}
  - \Mpl^{2} \Lambda b^{3}+\frac{ {\ddot{b}}^{2}  b\alpha_{CG}}{2}=0.
\end{equation}
The first solution is given by:
\begin{equation}
    b(t)=c_1\exp\left[\left(\frac{2\Lambda\Mpl^2}{\alpha_{CG}}\right)^{1/4}t\right]+c_2\exp\left[-\left(\frac{2\Lambda\Mpl^2}{\alpha_{CG}}\right)^{1/4}t\right],
\end{equation}
and characterizes the exponential expansion. The second solution also exists due to taking the square-root of the equation for the scale factor. It is given by:  
\begin{equation}
    b(t)=c_1\sin\left[\left(\frac{2\Lambda\Mpl^2}{\alpha_{CG}}\right)^{1/4}t\right]+c_2\cos\left[-\left(\frac{2\Lambda\Mpl^2}{\alpha_{CG}}\right)^{1/4}t\right]
\end{equation}
and it describes the closed universe in the anisotropic direction. 

One can easily verify that the above model does not admit power-law solutions, with $a\sim t^p$ and $b\sim t^n$.

\subsubsection{Numerical solutions}

As a next step, let us consider more general solutions to the background equations of motion. In order to find them, it is useful to notice that the system of background equations can be written in a simpler way. For this, we first consider the equations (\ref{5dconstraint}) and (\ref{5deoma}), and solve them for the fourth-order derivative of the scale factor $\ddddot{a}(t)$ and the cosmological constant $\Lambda$. Then, we substitute these expressions into the equation of motion for the other scale factor, given in (\ref{5deomb}). Upon this substitution, $\ddddot{b}(t)$ cancels, allowing us to solve the equation for $\dddot{b}(t)$. We then further use this expression to simplify the constraint equation, and the equation of motion for the other scale factor. The resulting system of equations is given by:
\begin{equation}\label{eqa}
    \begin{split}
      \dddot{b} &= 
\frac{1}{5 a^{3} b \left(b \dot{a}-\dot{b} a\right)}\left[5\dddot{a} a^{2} b^{3} \dot{a}-3 {\ddot{a}}^{2} a^{2} b^{3}+6\ddot{a} a b^{3} \dot{a}^{2}-8 b^{3} \dot{a}^{4}-5\dddot{a} \dot{b} a^{3} b^{2}+6\ddot{b}\ddot{a} a^{3} b^{2}-16\ddot{b} a^{2} b^{2} \dot{a}^{2}\right.\\&\left.-\ddot{a} \dot{b} a^{2} b^{2} \dot{a}+21 \dot{b} a b^{2} \dot{a}^{3}-3 {\ddot{b}}^{2} a^{4} b+21\ddot{b} \dot{b} a^{3} b \dot{a}-5\ddot{a} \dot{b}^{2} a^{3} b-18 \dot{b}^{2} a^{2} b \dot{a}^{2}-5\ddot{b} \dot{b}^{2} a^{4}+5 \dot{b}^{3} a^{3} \dot{a}\right]
    \end{split}
\end{equation}

\begin{equation}\label{eqb}
    \begin{split}
      \ddddot{a} &= 
\frac{1}{25 b^{2} a^{3} \left(\dot{b} a-b \dot{a}\right)^{2}}\left[\left(-3 b {\ddot{b}}^{3}-70 \dot{b}^{2} {\ddot{b}}^{2}\right) a^{6}-170 \dot{a}^{6} \dot{b}^{4}+\left(324 b^{4} \dot{a}^{4}\ddot{a}+914 b^{3} \dot{a}^{5} \dot{b}\right) a\right.\\&\left.-248 \dot{a}^{6} b^{4}+\left(9 b^{2}\ddot{a} {\ddot{b}}^{2}+\left(224 \dot{a} \dot{b} {\ddot{b}}^{2}-75 \dot{b}^{3}\dddot{a}+115 \dot{b}^{2}\ddot{a}\ddot{b}\right) b+240 \dot{a} \dot{b}^{3}\ddot{b}\right) a^{5}\right.\\&\left.+\left(-9 b^{3} {\ddot{a}}^{2}\ddot{b}+\left(-154 \dot{a}^{2} {\ddot{b}}^{2}+\left(225 \dot{b}^{2}\dddot{a}-398\ddot{a}\ddot{b} \dot{b}\right) \dot{a}-45 \dot{b}^{2} {\ddot{a}}^{2}\right) b^{2}+\left(-929 \dot{a}^{2} \dot{b}^{2}\ddot{b}-115 \dot{a} \dot{b}^{3}\ddot{a}\right) b\right) a^{4}\right.\\&\left.+\left(3 b^{4} {\ddot{a}}^{3}+\left(\left(-225 \dot{b}\dddot{a}+283\ddot{a}\ddot{b}\right) \dot{a}^{2}+174 \dot{a} \dot{b} {\ddot{a}}^{2}\right) b^{3}+\left(1138 \dot{a}^{3} \dot{b}\ddot{b}+554 \dot{a}^{2} \dot{b}^{2}\ddot{a}\right) b^{2}+758 b \dot{a}^{3} \dot{b}^{3}\right) a^{3}\right.\\&\left.+\left(\left(75 \dot{a}^{3}\dddot{a}-129 {\ddot{a}}^{2} \dot{a}^{2}\right) b^{4}+\left(-449 \dot{a}^{4}\ddot{b}-763\ddot{a} \dot{b} \dot{a}^{3}\right) b^{3}-1254 b^{2} \dot{a}^{4} \dot{b}^{2}\right) a^{2}\right]
    \end{split}
\end{equation}
and 
\begin{equation}\label{cc}
   \Lambda = 
\frac{\left(a^{2}\ddot{b}+\left(-\ddot{a} b-\dot{b} \dot{a}\right) a+b \dot{a}^{2}\right)^{2} \alpha_{CG}}{2 b^{2} a^{4} \Mpl^{2}}
\end{equation}

In the following, we will show two examples of the numerical solutions. In particular, we numerically solve the equations (\ref{eqa}) and (\ref{eqb}), and determine the value cosmological constant by using the constraint equation (\ref{cc}). In the plots below, we will set 
\begin{equation}
    \alpha_{CG}=1\qquad\text{and}\qquad  \Mpl=1. 
\end{equation}
The pink curve corresponds to the evolution of the scale factor $a(t)$, and the second, green one, corresponds to the scale factor $b(t)$. For each curve, we will present also the Hubble parameters, 
\begin{equation}
    H_a=\frac{\dot{a}}{a}\qquad\text{and}\qquad H_b=\frac{\dot{b}}{b}
\end{equation}
and the corresponding effective equations of state:
\begin{equation}
    \omega_a=-1-\frac{2}{3}\frac{\dot{H}_a}{H_a^2}\qquad\text{and}\qquad  \omega_b=-1-\frac{2}{3}\frac{\dot{H}_b}{H_b^2}. 
\end{equation}
Let us now consider two cases, depending on the initial conditions. One should note that the cases below are just examples, and one can find similar behavior for other choices of initial conditions as well. 
\begin{center}
    \textit{Case 1: Transition:  decelerated  -- super accelerated -- de Sitter expansion}
\end{center}
We first consider the possibility of super-Hubble evolution. By setting the initial conditions to 
\begin{equation}
    \begin{split}
        &a(0.1)=1\qquad\dot{a}(0.1)=0.2\qquad\ddot{a}(0.1)=-0.06\qquad\dddot{a}=0.2\\
        &b(0.1)=1\qquad \dot{b}(0.1)=0.8\qquad \ddot{b}(0.1)=0
    \end{split}
\end{equation}
we find: 
\begin{figure}[h]
    \centering

    \begin{minipage}{0.3\textwidth}
        \centering
        \includegraphics[width=\linewidth]{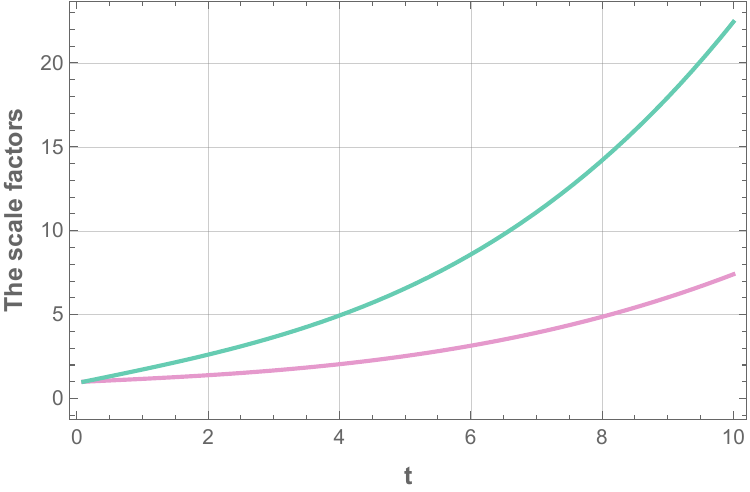}
    \end{minipage}
    \hfill
    \begin{minipage}{0.3\textwidth}
        \centering
        \includegraphics[width=\linewidth]{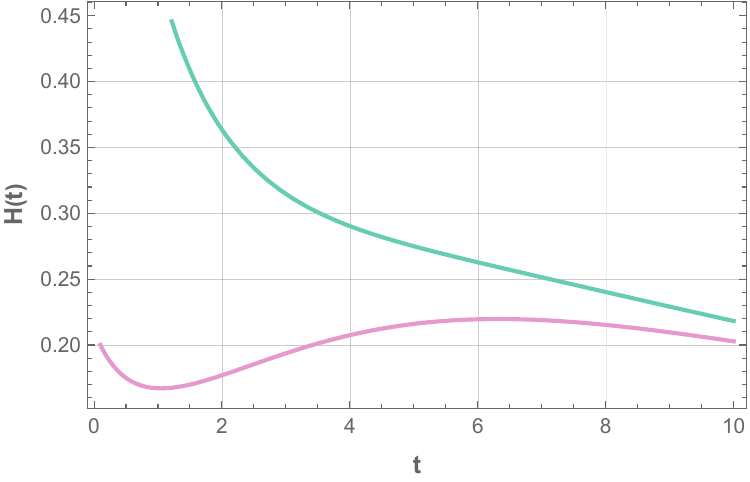}
    \end{minipage}
    \hfill
    \begin{minipage}{0.3\textwidth}
        \centering
        \includegraphics[width=\linewidth]{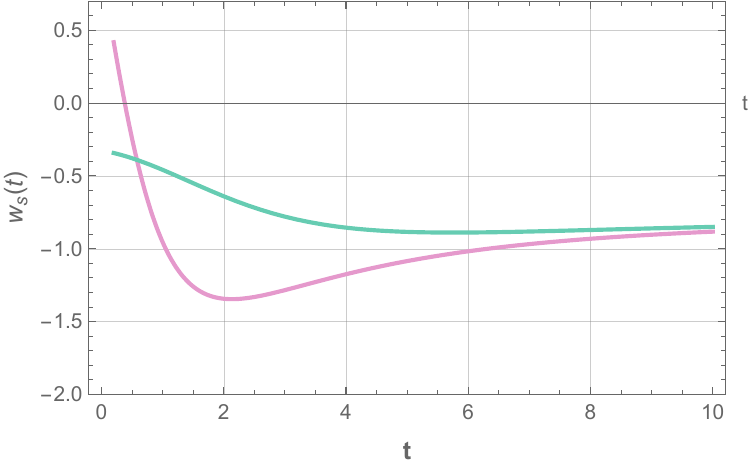}
    \end{minipage}

    \caption{Solution to the background equations of motion in the first case.  The green line corresponds to the anisotropic direction, characterized by the scale factor $b(t)$. The pink line corresponds to the scale factor of the isotropic subspace, $a(t)$. }

    \label{fig:threeinrow}
\end{figure}

\begin{center}
    \textit{Case 2: Transition:  accelerated expansion}
\end{center}
We next consider the case when the equation of state does not cross unity. For the following conditions, we find: 
\begin{equation}
    \begin{split}
        &a(0.1)=0.7\qquad\dot{a}(0.1)=0.7\qquad\ddot{a}(0.1)=0.2\qquad\dddot{a}=0\\
        &b(0.1)=0.5\qquad \dot{b}(0.1)=0.6\qquad \ddot{b}(0.1)=0.1
    \end{split}
\end{equation}
\begin{figure}[h]
    \centering

    \begin{minipage}{0.3\textwidth}
        \centering
        \includegraphics[width=\linewidth]{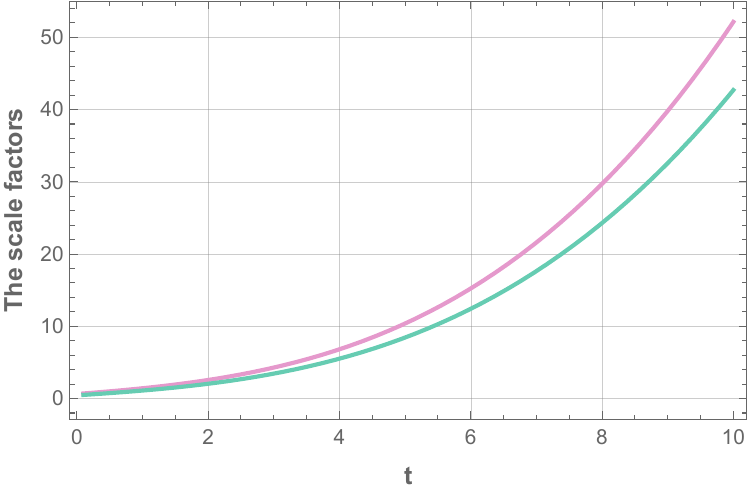}
    \end{minipage}
    \hfill
    \begin{minipage}{0.3\textwidth}
        \centering
        \includegraphics[width=\linewidth]{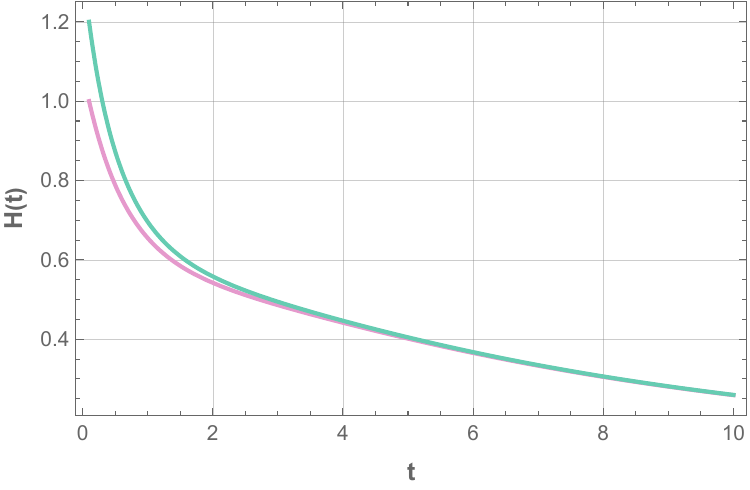}
    \end{minipage}
    \hfill
    \begin{minipage}{0.3\textwidth}
        \centering
        \includegraphics[width=\linewidth]{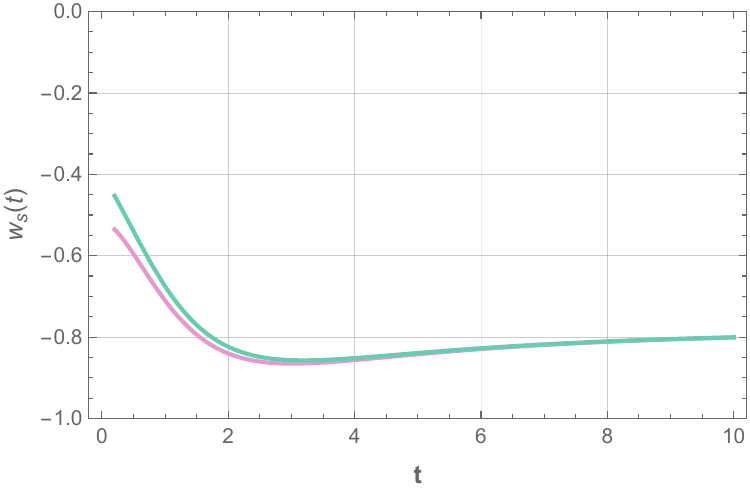}
    \end{minipage}

    \caption{Solution to the background equations of motion in the second case.  The green line again corresponds to the anisotropic direction, characterized by the scale factor $b(t)$. The pink line corresponds to the scale factor of the isotropic subspace, $a(t)$. }

    \label{fig:threeinrow}
\end{figure}

\subsection{Perturbations}
Let us now consider the perturbations around the previous background, similarly to the formalism developed in \cite{vandeBruck:2000ju}:
\begin{equation}
    g_{\mu\nu}= g_{\mu\nu}^{(0)}+\delta g_{\mu\nu}
\end{equation}
where 
\begin{equation}
    \begin{split}
        &\delta g_{00}=2\phi\\
        &\delta g_{oi}=S_i+B_{,i},\qquad S_{i,i}=0\\
         &\delta g_{0l}=\omega_{,l}\\
        &\delta g_{ij}=a^2(2\psi \delta_{ij}+2E_{,ij}+F_{i,j}+F_{j,i}+h_{ij}^T), \qquad F_{i,i}=0,\qquad h^T_{ii}=0,\qquad h^T_{ij,i}=0\\
         &\delta g_{ll}=b^2\sigma\\
        &\delta g_{li}=v_{i,l}+\mu_{,li}\qquad v_{i,i}=0
    \end{split}
\end{equation}
Here, $i$ runs over $x,y,$ and $z$ coordinates. Under the infinitesimal transformations, 
\begin{equation}
    x^{\mu}\to\Tilde{x}^{\mu}=x^{\mu}+\xi^{\mu},
\end{equation}
the above components transform as:
\begin{equation}
    \begin{split}
        \Tilde{\phi}&=\phi+\dot{\xi}^{0}\\
        \Tilde{B}&=B+\xi^0-a^2\dot{\zeta}\\
        \Tilde{\omega}&=\omega+\xi^{0}-b^2\dot{\chi}\\
        \Tilde{\psi}&=\psi-\frac{\dot{a}}{a}\xi^0\\
        \Tilde{E}&=E-\zeta\\
        \Tilde{\mu}&=\mu-b^2\chi-a^2\zeta\\
        \Tilde{S}_i&=S_i-a^2\dot{\xi}_i^T\\
        \Tilde{F}_i&=F_i-\xi_i^T\\
        \Tilde{v}_i&=v_i-a^2\xi_i^T\\
        \tilde{h}_{ij}^T&=h_{ij}^T
    \end{split}
\end{equation}
In the above, we have decomposed $\xi^{\mu}$ as:
\begin{equation}
    \xi^{0}\qquad \xi_i=\delta_{ij}\xi^j=\xi_i^T+\zeta_{,i}\qquad \xi^l=\chi_{,l}
\end{equation}
with $\xi_{i,i}^T=0.$
We can see that the tensor perturbations are gauge invariant. Among the vector perturbations, only two are independent. Thus, we can form two gauge invariant variables:
\begin{equation}
    \Bar{V}_i=S_i-a^2\dot{F}_i\qquad\text{and}\qquad \Bar{U}_i=v_i-a^2F_i. 
\end{equation}
Finally, among the scalar modes, three are dependent. Similarly to the cosmological perturbations in four dimensions \cite{Bardeen:1980kt, Mukhanov:1990me}, we can form the following gauge invariant variables: 
\begin{equation}
    \begin{split}
        \Phi&=\phi-\left(B-a^2\dot{E}\right)_{,0}\\
        \Psi&=\psi +\frac{\dot{a}}{a}\left(B-a^2\dot{E}\right)\\
        \Omega&=\omega-\frac{\dot{a}}{a}\left(B-a^2\dot{E}\right)-\left(\frac{1}{b^2}\mu-\frac{a^2}{b^2}E\right)_{,0}
    \end{split}
\end{equation}
Therefore, in the following, we will work in the gauges: 
\begin{equation}
    E=0\qquad B=0\qquad\mu=0\qquad F_i=0, 
\end{equation}
in which the variables coincide with the gauge invariant ones, 
and separately analyze scalar, vector and tensor perturbations. One should note that due to the length of the expressions, we will avoid writing any explicit coefficients in the following, but rather write the important structure of the equations, as well as the procedure. 

\subsubsection{The scalar perturbations}
Let us first consider the scalar perturbations. For this, we expand the action up to second order in perturbations, setting the vector and tensor ones to zero. During the computation, we will always assume that the background equations of motion (\ref{eqa}), (\ref{eqb}) and (\ref{cc})  are satisfied. It is convenient to analyze the resulting expressions in terms of the  Fourier modes:
\begin{equation}
    X=\int \frac{d^3kdq}{(2\pi)^{3/2}}X(t,k,q)e^{ik_ix^i+iql}, \qquad x^i={x,y,z}
\end{equation}
where $X$ stands for all of the scalar modes. First, we perform several integrations by parts to remove all of the fourth- and third-order derivatives from the modes. Then, we can notice that by substituting 
\begin{equation}
    \sigma=\sigma_2+\psi
\end{equation}
and performing integration by parts, among all of the modes, only  $\sigma_2$ appears higher time derivatives, in the form $\ddot{\sigma}_2\ddot{\sigma}_2$, while the other fields are coupled to it through terms such as $\ddot{\sigma}_2\dot{X}$, with $X=\{\psi_2,\omega,\phi, \sigma_2\}$.  In addition, the remaining modes have standard kinetic terms in the Lagrangian, with the form $\dot{X}\dot{X}$.  By substituting
\begin{equation}
    \psi=\psi_2+\phi,
\end{equation}
one can notice that $\phi$ loses its kinetic term. By varying the action with respect to it, we find its corresponding constraint, which we solve and substitute back to the action. The resulting action is a function of three fields, $\{\psi_2,\omega, \sigma_2\}$, all of which are propagating. In particular, the Lagrangian of the system is given in the following form: 
\begin{equation}
    \begin{split}
\mathcal{L}&=a_1\ddot{\sigma}_2\ddot{\sigma}_2+a_2\ddot{\sigma}_2\dot{\psi}_2+a_3\ddot{\sigma}_2\dot{\omega}+a_4\dot{\sigma}_2\dot{\sigma}_2+a_5\dot{\psi}_2\dot{\psi}_2+a_6\dot{\omega}\dot{\omega}\\&+a_7\dot{\psi}_2\dot{\omega}+a_7\dot{\psi}_2\dot{\sigma}_2+a_8\dot{\sigma}_2\dot{\omega}+a_9\dot{\psi}_2\sigma_2+a_{10}\dot{\psi}_2\omega+a_{11}\dot{\sigma}_2\omega\\&
+a_{12}\psi_2\sigma_2+a_{13}\psi_2\omega+a_{14}\omega\sigma_2+a_{15}\psi_2^2+a_{16}\sigma_2^2+a_{17}\omega_2^2
    \end{split}
\end{equation}
where $a_1 ... a_{17}$ are coefficients which depend on time and momentum $k$ and $q$. One should note that each of these modes depends on the momentums. Thus, the above should be read as: 
\begin{equation}
    a XY\equiv a(t,k,q)\left(X(t,k,q)Y(t,-k,-q)+X(t,-k,-q)Y(t,k,q)\right)
\end{equation}
where $X$ and $Y$ represent the above fields and their time derivatives. 
On the first sight, this seems like a system that describes four degrees of freedom. The first field appears with fourth order time derivatives, while the second two with first order terms. However, as pointed out in \cite{Langlois:2015skt}, as the three fields are coupled, it may happen that the whole system is degenerate, meaning that not all equations are independent. This is what happens in the degenerate higher-order scalar-tensor theories (DHOST) \cite{BenAchour:2016cay, Langlois:2017mdk}. To find out if this is the case, we reduce the previous system to a first-order one, by defining 
\begin{equation}
    \rho=\dot{\sigma},
\end{equation}
implementing it in the above Lagrangian with the Lagrangian multiplier, and treating the variables,
\begin{equation}
    {\rho, \sigma_2, \psi_2, \omega}
\end{equation}
as independent fields. Then, we can easily check that the kinetic matrix associated with the fields $\rho, \psi_2, \omega$ is degenerate, meaning that its determinant vanishes. Thus, the above system is not fully independent, but rather has a constraint that reduces the overall number of degrees of freedom by one. As an example, we can demonstrate this on the level of the equations of motion, with an analytical solution that we have previously found, corresponding to $b(t)=1$. By varying the action with respect to $\psi_2$, $\omega$ and $\sigma_2$ we obtain respectively the equations of motion which the form: 
\begin{equation}\label{eqPs}
b_1\ddot{\psi}_2+b_2\dddot{\sigma}_2+b_3\ddot{\sigma}_2+b_4\dot{\psi}_2+b_5\dot{\sigma}_2+b_6\psi_2+b_7\sigma_2+b_8\ddot{\omega}+b_9\dot{\omega}+b_{10}\omega=0,
\end{equation}
\begin{equation}\label{eqOm}
d_1\ddot{\omega}+d_2\dddot{\sigma}_2+d_3\ddot{\sigma}_2+d_4\dot{\omega}+d_5\dot{\sigma}_2+d_6\omega+d_7\sigma_2+d_8\ddot{\psi}_2+d_9\dot{\psi}_2+d_{10}\psi_2=0
\end{equation}
and 
\begin{equation}\label{eqSi}
c_1\ddddot{\sigma}_2+c_2\dddot{\sigma}_2+c_3\dddot{\psi}_2+c_4\ddot{\sigma}_2+c_5\ddot{\psi}_2+c_6\dot{\sigma}_2+c_7\dot{\psi}_2+c_8\sigma_2+c_9\psi_2+c_{10}\ddot{\omega}+c_{11}\dot{\omega}+c_{13}{\omega}+c_{14}\dddot{\omega}=0
\end{equation}
In the above, all fields are functions of the two momenta and time $(t,k,q)$, and we have not performed any approximation in high k or high q limits. The coefficients $b_1...b_{10}$, $d_1...d_{10}$ and $c_1 ... c_{14}$ are functions of time and momenta. 
By further substituting 
\begin{equation}
    \psi_2=\psi_3+\frac{ab}{\dot{a}b-\dot{b}a}\dot{\sigma}_2
\end{equation}
into the above equations, we find that in (\ref{eqPs}) and (\ref{eqOm}), $c_1\ddddot{\sigma}_2$ cancels, making the equation second-order in the time derivatives. Moreover, equation (\ref{eqSi}) becomes third-order differential equation, with fourth-order derivatives cancelling. By taking a time derivative of (\ref{eqPs}), we can use the resulting expression, and cancel the third-order derivatives in (\ref{eqSi}). Then, we obtain a system of three second order differential equations: 
\begin{equation}
    A(t)\ddot{V}+B(t)\dot{V}+C(t)=0
\end{equation}
where 
\begin{equation}
    V=\left[\begin{array}{c}
{\psi_3} \! \left(t , k , q\right) 
\\
 {\sigma_2} \! \left(t , k , q\right) 
 \\
  {\omega} \! \left(t , k , q\right) 
\end{array}\right]
\end{equation}
Here, $A$, $B$, $C$ are the associated matrices for the system, that depend on the scale factor as well as the momentas. In particular, the matrix $A$, is given by:
\begin{equation}
  A= \alpha_{CG} \left[\begin{array}{ccc}
-\frac{30 \Mpl \,\dot{b}^{2}  }{11 b} & \frac{10 \Mpl \left(k^{2} b^{2}-73 \dot{b}^{2}-3 q^{2}\right)  }{11 b} & -\frac{30 \Mpl    \,q^{2} \dot{b}}{11 b^{2}} 
\\
A_{21} &A_{22} & A_{23}
\\
 -\frac{30 \Mpl    \,q^{2} \dot{b}}{11 b^{2}} & -\frac{10 \Mpl \,q^{2} \left(19 k^{2} b^{2}+219 \dot{b}^{2}+9 q^{2}\right)  }{33 b^{2} \dot{b}} & -\frac{10 \Mpl \,q^{2} \left(k^{2} b^{2}+\frac{9 q^{2}}{11}\right)  }{3 b^{3}} 
\end{array}\right]
\end{equation}
with
\begin{equation}
 \begin{split}
  A_{21}&=  \alpha_{CG}\frac{100 \Mpl^{2}}{121 b^{2}} \left(k^{2} b^{2}-73 \dot{b}^{2}-3 q^{2}\right) \left(k^{2} b^{2}-3 q^{2}\right)\\
        A_{22}&= -\alpha_{CG}\frac{100 \Mpl^{2}}{131769 b^{2} \dot{b}^{2}}  \left(k^{2} b^{2}-73 \dot{b}^{2}-3 q^{2}\right) \times \\&\times\left(5687 k^{4} b^{4}+93621 k^{2} \dot{b}^{2} b^{2}+13794 k^{2} q^{2} b^{2}+232246560 \dot{b}^{4}-280863 \dot{b}^{2} q^{2}+3267 q^{4}\right)\\
       A_{23}&=-\alpha_{CG}\frac{1900 \Mpl^{2} q^{2}}{363 b^{3} \dot{b}}  \left(k^{2} b^{2}+\frac{9 q^{2}}{19}\right) \left(k^{2} b^{2}-73 \dot{b}^{2}-3 q^{2}\right) 
 \end{split}
\end{equation}
We can see that its determinant is not vanishing, and thus indicates that the above system of equations is not degenerate. Therefore, overall, in the scalar sector, we find three degrees of freedom.

\subsubsection{Vector perturbations}
Similarly to the scalar perturbations, we expand the action up to second order in the vector perturbations. We can express them in terms of the Fourier modes as:
\begin{equation}
    X_i=\int \frac{d^3kdq}{(2\pi)^{3/2}}\varepsilon_iX(t,k,q)e^{ik_ix^i+iql}, \qquad x^i={x,y,z}
\end{equation}
where $\varepsilon_i$ are the polarization vectors associated with the homogeneous and isotropic subspace, and assume that the background equations of motion are satisfied. We integrate by parts, so that the fourth-order and cubic-order derivatives dissapear from the Lagrangian. Then, we find the Lagrangian density in the following form:  
\begin{equation}
\mathcal{L}=f_1\ddot{v}_i\ddot{v}_i+f_2\ddot{v}_i\dot{S}_i+f_3\dot{v}_i\dot{v}_i+f_4\dot{S}_i\dot{S}_i+f_5\dot{V}_i\dot{S}_i+f_6S_iS_i+f_7v_iv_i+f_7S_iv_i
\end{equation}
Here, $f_1 ... f_7$ are time and momentum dependent functions. 
We can notice that similarly to the scalar sector, due to the anisotropy in the fifth dimension, the the vector modes $v_i$ acquire higher-order term in the derivatives. However, this does not imply that the total number of degrees of freedom is three (per index i), due to the $f_2$ term. In particular, one can show that this theory is degenerate as well -- by reducing it to the system of first order equations with introduction of $\rho_i=\dot{v}_i$, we find that the determinant of the corresponding kinetic matrix is not zero. Thus, overall the total number of vector modes is 3 dof, per index $i$.

\subsubsection{Tensor perturbations}
Finally, let us consider the tensor perturbations. We again expand the action up to second oder in the perturbations. In the Fourier space, we then further express the tensor modes as: 
\begin{equation}
    h_{ij}^T=\int \frac{d^3kdq}{(2\pi)^{3/2}}\varepsilon_{ij}^TX(t,k,q)e^{ik_ix^i+iql}, \qquad x^i={x,y,z}
\end{equation}
where $\varepsilon_{ij}^T$ are the corresponding polarization tensors associated with the homogeneous and isotropic subspace. Assuming that the background equations of motion are satisfied, we then find the Lagrangian density of the following form: 
\begin{equation}
\mathcal{L}=n_1\ddot{h}_{ij}^T\ddot{h}_{ij}^T+n_2\ddot{h}_{ij}^T\dot{h}_{ij}^T+n_3\dot{h}_{ij}^T\dot{h}_{ij}^T+n_4\dot{h}_{ij}^Th_{ij}^T+n_5h_{ij}^Th_{ij}^T
\end{equation}
Here, $n_1 ... n_5$ are time and momentum dependent coefficients. In contrast to scalar and vector perturbations, there is only one type of tensor modes, which is propagating. 
Therefore, similarly to the conformal gravity in four dimensions, this theory has two tensor modes, among which one is a ghost.

\section{Discussion and summary}

Higher-derivative theories of gravity have always taken an important role in modified theories of gravity, as they provide an example of renormalizable theories in four dimensions. At the same time, such theories also arise in extra-dimensional scenarios, thus making it important to understand their structure. In this work, we have studied these theories in higher dimensions, with a special focus on pure scale-invariant gravity, and conformal gravity. 

First, we have studied pure scale-invariant gravity in d dimensions, which contains only powers of the Ricci tensor. We have seen that this theory shares a lot of similarities with the four-dimensional case. In particular, we have shown that even though the first-order corrections are non-linear in the Jordan frame for the flat-spacetime background, there, the theory does not propagate any degrees of freedom. In contrast, for $R\neq 0$, the theory propagates a scalar mode and two tensor ones, which can be easily seen in the Einstein frame. However, in the four-dimensional case, the theory contains a cosmological constant that can be either positive or negative in the Einstein frame, for higher dimensions, the cosmological constant can only take positive values, due to the powers of the coupling constant of the theory.  

As a second interesting model, we have considered conformal gravity in d dimensions, whose action was built from the square of the Weyl tensor: $W_{\mu\nu\rho\sigma}W^{\mu\nu\rho\sigma}$. First, we have studied this theory for conformally flat space-time. Unfortunately, due to the non-linear nature, in this case, the scalar, vector, and tensor modes do not decouple, and this makes it difficult to write the action only in terms of the propagating degrees of freedom. However, by considering the structure of the constraints and the equations of motion, it is clear that the theory will propagate a vector mode, and two tensor ones, in agreement with the four-dimensional case. 

In order to further explore the field content of conformal gravity, we have found a frame, in which the action becomes linear in $W_{\mu\nu\rho\sigma}W^{\mu\nu\rho\sigma}$, and coupled to the cosmological constant. Notably, in this frame, all fields are invariant under conformal transformations, making it thus exciting to apply boundary conditions without explicitly breaking conformal invariance, but removing the ghost degrees of freedom, along the approach of \cite{Maldacena:2011mk, Hell:2023rbf}. The new frame can also be extended to $f(W^2)$ theory. In particular, we have found that in this case, the cosmological constant becomes replaced with a scalar field, which, in contrast to the $f(R)$ gravity, is not propagating. 

In the last section, we have focused on the particular model in five dimensions and analyzed background solutions and the number of propagating modes for the theory in the alternative frame. This formulation exists for backgrounds that are not conformally flat. Thus, as the simplest case, we have considered an anisotropic background. By studying the background solutions, we have found that the theory admits at least two interesting analytic ones. The first allows for the unit scale factor in the anisotropic direction, while the homogeneous subspace expands in a super-Hubble way. In the other case, the homogeneous subspace is flat, while the anisotropic direction admits solutions corresponding to a closed universe or exponential expansion. In addition, we have studied the background solutions numerically. There, we have found two interesting cases. In the first case, the homogeneous subspace expands exponentially. The universe in the anisotropic direction, on the other hand first decelerates, then transitions to a super-accelerated phase, and then later transitions to the de-sitter expansion. Interestingly, the equation of state corresponding to this direction is phantom-like, with $\omega_b<-1$. In the second case, the two branches are accelerating.

Finally, we have studied the number of modes for these equations. We have seen that in the tensor sector, the theory matches with the four-dimensional case -- it propagates two tensor modes, one of which is healthy, and one which is a ghost dof. Notably, we have also seen that the theory is degenerate in the scalar sector. In particular, at first sight, it may appear that the theory propagates four scalar modes. However, the kinetic matrix associated with the system indicates that one of them is redundant, as its determinant is not vanishing. By verifying this further on the level of equations of motion, we have confirmed that there are three scalar modes in total. The number of vector modes increases as well, having now three modes propagating per index i. These scalar and vector modes do not appear in the four-dimensional case, thus increasing the spectrum in the presence of extra dimensions. 

Overall thus, we find that higher-derivative theories of gravity may allow for a larger spectrum of modes when compared to their four-dimensional analogues. It will be thus interesting to explore if this could change by the addition of other terms, or if the ghost dof could be removed in the alternative frame with boundary conditions. 

\begin{center}

\textbf{Acknowledgments}
\end{center}

\textit{A. H. would like to thank R. Brandenberger and Elisa G. M. Ferreira for a nice discussion. A. H. also thanks LMU Munich, ASC center and the RBI Zagreb for hospitality during her visit, when part of her work was performed. The work of A. H. was supported
by the World Premier International Research Center Initiative (WPI), MEXT, Japan. D.L. thanks the IPMU for hospitality during his visit, when part of this work was performed.
The work of D.L. is supported by the Origins Excellence Cluster and by the German-Israel-Project (DIP) on Holography and the Swampland.}

\bibliographystyle{utphys}
\bibliography{paper}

\end{document}